\documentclass[apj,numberedappendix]{emulateapj}
\usepackage{apjfonts}

\shorttitle{PARTIALLY ECLIPSING WD BINARY J1435}
\shortauthors{STEINFADT, BILDSTEN, \& HOWELL}

\newcommand{\be}{\begin{eqnarray}}
\newcommand{\ee}{\end{eqnarray}}

\newcommand{\scaleeps}{\epsscale{1.0}}

\begin{document}


\title{Discovery of the Partially Eclipsing White Dwarf Binary SDSS J143547.87+373338.5\altaffilmark{\textdagger}}

\author{Justin D. R. Steinfadt\altaffilmark{1,4}}

\author{Lars Bildsten\altaffilmark{1,2}}

\author{Steve B. Howell\altaffilmark{3,4}}

\altaffiltext{\textdagger}{The WIYN Observatory is a joint facility of the University of Wisconsin-Madison, Indiana University, Yale University, and the National Optical Astronomy Observatories.}
\altaffiltext{1}{Department of Physics, Broida Hall,\\University of California, Santa Barbara,  CA 93106,\\jdrs@physics.ucsb.edu}
\altaffiltext{2}{Kavli Institute for Theoretical Physics, Kohn Hall,\\University of California, Santa Barbara,  CA 93106,\\bildsten@kitp.ucsb.edu}
\altaffiltext{3}{WIYN Observatory and National Optical Astronomy Observatory,\\950 N. Cherry Ave., Tucson,  AZ 85719,\\howell@noao.edu}
\altaffiltext{4}{Visiting Astronomer, Kitt Peak National Observatory, National Optical Astronomy Observatories, which is operated by the Association of Universities for Research in Astronomy, Inc. (AURA) under cooperative agreement with the National Science Foundation.}


\begin{abstract}

We have discovered a partially eclipsing white dwarf, low-mass M dwarf binary (3.015114 hour orbital period), SDSS J143547.87+373338.5, from May 2007 observations at the WIYN telescope.  Here we present blue band photometry of three eclipses.  Eclipse fitting gives main sequence solutions to the M dwarf companion of $M_S=0.15-0.35 M_{\odot}$ and $R_S=0.17-0.32 R_{\odot}$.  Analysis of the SDSS spectrum constrains the M dwarf further to be of type M4-M6 with $M_S=0.11-0.20 M_{\odot}$.  Once full radial velocity curves are measured, high precision determinations of the masses and radii of both components will be easily obtained without any knowledge of stellar structure or evolution.  ZZ Ceti pulsations from the white dwarf were not found at our 4 mmag detection limit.

\end{abstract}



\section{Introduction}

Eclipsing binary systems offer unique opportunities for precise measurements of stellar properties.  With full radial velocity and light curves (commonly from eclipsing double-lined spectroscopic binaries), the masses and radii of both components are measured to very high precision and accuracy.

We discovered that the $u' = 17.7$ and $g' = 17.1$ mag binary SDSS J143547.87+373338.5 (hereafter J1435, \citealt{dje06}) undergoes a partial eclipse of the WD by its low-mass companion with a period of $3 \ \rm hrs$ and a transit time of $\approx 480 \ \rm sec$.  The eclipsing M dwarf secondary is expected to contribute $\lesssim 4\%$ in the blue band BG-39 filter.  Measurements by G\"{a}nsicke et al. (private communication, \citealt{arm07}) give an accepted range of WD masses of $0.35-0.58 M_{\odot}$ (see Table \ref{tbl:prop} and \S\ref{sec:ecl}).  Our analysis in \S\ref{sec:ecl} using model fits to the partial eclipse produce main sequence (MS) solutions using \citet{ib98} for the M dwarf of $M_S\approx0.15-0.35 M_{\odot}$ and $R_S\approx0.17-0.32 R_{\odot}$, making it a candidate for probing the stellar structure of the low-mass MS.  The M dwarf is further constrained in \S\ref{sec:dis} by using the SDSS spectrum and is expected to be of type M4-M6 with $M_S\approx0.11-0.20 M_{\odot}$.  Measuring the ingress and egress as well as ellipsoidal and reflection effects in multiple color filters of the J1435 eclipse would probe the WD atmosphere.

In \S\ref{sec:obs} we detail our observations and data reduction.  A deep analysis for ZZ Ceti type pulsations revealed no pulsations down to a 4 mmag detection limit.   In \S\ref{sec:ecl} we discuss our analysis of the eclipse light curve and use it to constrain the properties of J1435.  In \S\ref{sec:dis} we discuss the consequences our results along with the SDSS spectrum have on the properties of the secondary star.


\section{Observations and Data Reduction}
\label{sec:obs}

\begin{deluxetable}{lccc}
\tablewidth{0pt}
\tablecaption{Properties of SDSS J143547.87+373338.5\label{tbl:prop}}
\tablehead{
\colhead{Property} & \colhead{Eisenstein et} & \colhead{G\"{a}nsicke et} & \colhead{G\"{a}nsicke et} \\
\colhead{ } & \colhead{al. (2006)} & \colhead{al. (Hot)} & \colhead{al. (Cold)}  }
\tablecolumns{4}
\scriptsize
\startdata
	$T_{\rm eff}$ (K) & $11062 \pm 58$ & $12681 \pm  990$ & $12392 \pm  350$\\
	$\log g$ (dex) & $6.860 \pm 0.038$ & $7.68 \pm 0.20$ & $7.82 \pm 0.14$\\
	$M_{WD} \; (M_\odot)$ & $0.20$\tablenotemark{a} & $0.35$\tablenotemark{b} & $0.58$\tablenotemark{c}\\
	$R_{WD} \; (R_\odot)$ & $0.0273$\tablenotemark{a} & $0.0178$\tablenotemark{b} & $0.0132$\tablenotemark{c}\\
\enddata
\tablenotetext{a}{From \citet{lga97} He core model, using $\log g = 6.86$.}
\tablenotetext{b}{From \citet{lga97} He core model, using $\log g = 7.48$ lower limit.}
\tablenotetext{c}{From \citet{lga98} C/O core with $10^{-4} M_{\odot}$ H envelope model, using $\log g = 7.96$ upper limit.}
\end{deluxetable}

SDSS J1435 was targeted, along with many other objects, in a campaign to discover very-low-mass (He core) and very-high-mass (O-Ne core) pulsating DA WDs (ZZ Cetis).  From within the Sloan Digital Sky Survey, \citet{dje06} released over 9,000 spectroscopically classified WDs.  We selected only those DA objects within the empirical ZZ Ceti instability strip \citep{ag06,asm04} of either low ($\log \; g<7.45$) or high ($\log \; g>8.7$) gravity.  J1435 was among several dozen objects that met the low gravity criteria and was visible during our observing run of 29 May through 1 June 2007.  In Table \ref{tbl:prop} we list the properties of J1435 as measured by \citet{dje06} and G\"{a}nsicke et al. (priv. com., \citealt{arm07}).  We report both a "cold" and "hot" solution for the WD parameters as the equivalent widths of the Balmer lines go through a maximum near $T_{\rm eff}\approx13,000$ K (dependent on $\log g$) and two solutions of similar quality result on either side of this maximum.  In most cases, this degeneracy is lifted by fitting the overall shape of the spectrum, however, since J1435's $T_{\rm eff}$ is so near the maximum, both solutions are indistinguishable \citep{arm07}.  The large discrepancy in parameters is because \citet{dje06} did not spectrally subtract the companion's contribution whereas G\"{a}nsicke et al. (priv. com., \citealt{arm07}) did.


\begin{figure}
	\centering
	\scaleeps
	\plotone{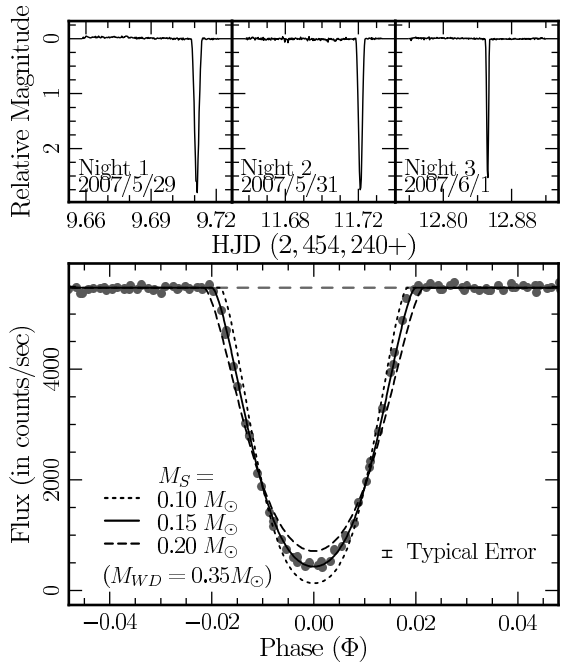}
	\caption{Top Panels: Light curves of all three nights in relative magnitudes.  Bottom Panel: Phased light curves of all nights centered on the primary eclipse and in relative flux units. Note the representative average error bar of the measurements (grey points). The solid line is the eclipse fit (see \S\ref{sec:ecl}) for $M_{WD}=0.35 M_{\odot}$ (first line in Table \ref{tbl:mss}).  The other two lines are the best possible fits for other MS masses at $M_{WD}=0.35 M_{\odot}$.}
	\label{fig:lc}
\end{figure}

We observed J1435 on the nights of 29 May, 31 May, and 1 June 2007 with the 3.5-meter WIYN telescope at the Kitt Peak National Observatory.  All observations used the OPTIC camera with two 4K$\times$2K pixel CCDs side-by-side ($15-\mu \rm m$ pixels) for a total of 4K$\times$4K pixel viewing area with a full frame field of view of 9.5$\times$9.5 arcminutes \citep{sbh03}.  Exposure times were 15 seconds over $\approx 2$ hr for the 29 May and 31 May observations, and 40 seconds over $\approx 3.5$ hrs on 1 June.  All exposures were with the broadband BG-39 filter ($\lambda_c \approx 4800$\AA, $\rm{FWHM} \approx 2600$\AA).  The 4K$\times$4K total pixel CCD was binned 2$\times$2 to reduce the readout time to $\approx 8.1$ seconds, which varied by $\approx0.03$ seconds during a night.

All images were reduced by applying an averaged two dimensional bias subtraction, and were flattened using averaged and normalized dome flat field images taken immediately prior to the night's observations.  All tasks were performed using standard tasks within IRAF\footnote{IRAF (Image Reduction and Analysis Facility) is distributed by the National Optical Astronomy Observatory, which is operated by the Association of Universities for Research in Astronomy, Inc., under contract with the National Science Foundation.  http://iraf.noao.edu}.  Light curves were produced by extracting fluxes from the program star and several comparison stars and taking their ratios.  Fluxes were extracted using the IRAF package VAPHOT \citep{hjd01}, which applies a dynamic method to aperture size adjustment (both flux apertures and sky background estimation annuli) to account for variable seeing.  Fifteen comparison stars were selected within the frame ranging from \textit{B} $\approx$ 16 to 19.5.  Differential photometry was performed using the comparison star weighting scheme detailed in \citet{jls01} and inspired by \citet{rlg88}.

Though targeted as a possible ZZ Ceti using the $T_{\rm eff}$ and $\log g$ measured by \citet{dje06}, the new $T_{\rm eff}$ and $\log g$ measurements by G\"{a}nsicke et al. (priv. com., \citealt{arm07}) place it outside the empirical instability strip \citep{ag06,asm04}.  It is therefore not surprising that our  Lomb-Scargle periodogram analysis found no (to 4 mmags) pulsations in the frequency range of 1-17 mHz.


\section{Eclipse Analysis}
\label{sec:ecl}

All three nights of observation show distinct primary eclipses (see Figure \ref{fig:lc}).  Using the technique developed by \citet{kkk56}, fits to the transit centers of these eclipses show primary minima at HJD $2,454,249.711028 \pm 0.000016, \; 2,454,251.7211338 \pm 0.0000044, \;$ and $2,454,252.8517916 \pm 0.0000051$.  Combining these primary minimum epochs gives an ephemeris of HJD $2,454,249.711056 \pm 0.000011 \; + \; \rm{N} \;\; 0.12562974 \pm 0.00000055$.  Our analysis obtains an orbital period of $P=3.015114 \pm 0.000013$ hrs ($10,854.410 \pm 0.048$ sec) and a transit time (measured as first departure from and full recovery to maximum light) of $\tau \approx 480$ sec.

We constrain the secondary through detailed modeling of the eclipse light curve.  There is no evidence of a secondary eclipse, so we only model the primary eclipse and assume the secondary star to be a black (non-luminous) disk.  If we assume a point light source WD, $R_{WD} \ll R_S$, Kepler's Law and the system's partially eclipsing geometry lead to a degeneracy between $M_S$, $R_S$, and $i$.  We can connect $M_S$ and $R_S$ with a MS $M-R$ relation, but are unable to break the relationship to $i$.  This last degeneracy is broken by giving the WD its finite physical extent, $R_{WD}$, and modeling the WD disk with a standard linear limb darkening law \citep{wvh93}:  $I(\theta)/I(0^{\circ}) = 1-u_{LD} \left[ 1-\cos(\theta) \right]$, where $\theta$ is the angle made by the normal vector to the WD surface to the line-of-sight.  We obtain the range $M_{WD}=0.35-0.58 M_{\odot}$ by using the $\log g$ and $T_{\rm eff}$ ranges measured by G\"{a}nsicke et al. (priv. com., \citealt{arm07}, see Table \ref{tbl:prop}).  The low mass limit uses the He core WD cooling tables of \citet{lga97} while the high mass limit uses the cooling tables of \citet{lga98} for a C/O core WD with a $10^{-4} M_{\odot}$ H envelope of solar metallicity.  Finally, we assume circular orbits and the geometry of an inclined system of orbiting masses.  In all, there are seven parameters: $M_{WD}$, $M_{S}$, $R_{WD}$, $R_{S}$, $i$, $P$, and $u_{LD}$.  We fit simple $\cos(\Phi)$ and $\cos(2 \Phi)$ functions to determine the strength of reflection and ellipsoidal variations (respectively) on the light curve, but found no such variations at the 0.5\% level.  We neglect the modeling of these variations.

\begin{figure}
	\centering
	\scaleeps
	\plotone{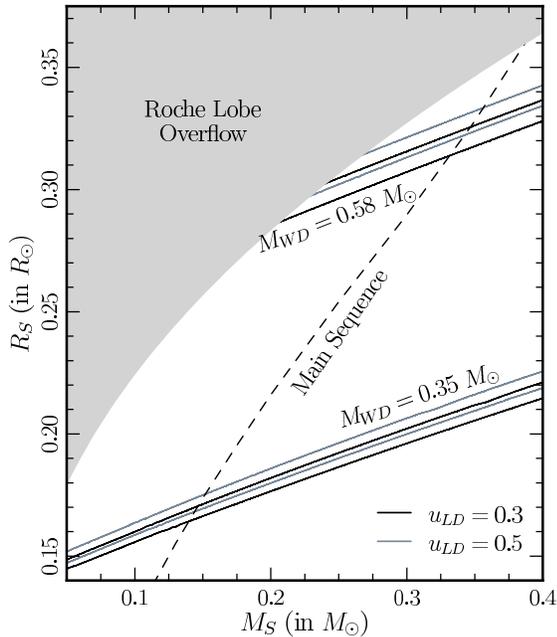}
	\caption{Constraints within the $R_S-M_S$ parameter space. Three-sigma contours of minimized $\chi^2_{\rm fit}$ to eclipse light curve (see \S\ref{sec:ecl}) using the two extreme accepted WD masses and two linear limb darkening coefficients. RL overflow boundary defined using \citet{ppe83}. Theoretical MS from \citet{ib98}.}
	\label{fig:rmch}
\end{figure}

In Figure \ref{fig:rmch} we present the three-sigma $\chi^2_{\rm fit}$ contours for the $R_S-M_S$ parameter space for two values of the linear limb darkening coefficient and extreme WD masses of the accepted range.  We obtain a contour by fixing $P$, $u_{LD}$, $M_{WD}$, and $R_{WD}$.  Then at each point in the $R_S-M_S$ parameter space we minimize $\chi^2_{\rm fit}$ with respect to the inclination and adopt this $\chi^2_{\rm fit}$ value for that mass and radius.  Our data tightly constrains the radius of the secondary while very loosely constraining the mass if we assume a specific linear limb darkening coefficient and WD solution with non-negligible radius.  Figure \ref{fig:lc} shows a comparison of model fits to the observed light curve for one set of WD parameters and three sets of MS parameters fit over $i$.  This figure plainly shows the broken degeneracy between $M_S-R_S$ and $i$.

\begin{figure}
	\centering
	\scaleeps
	\plotone{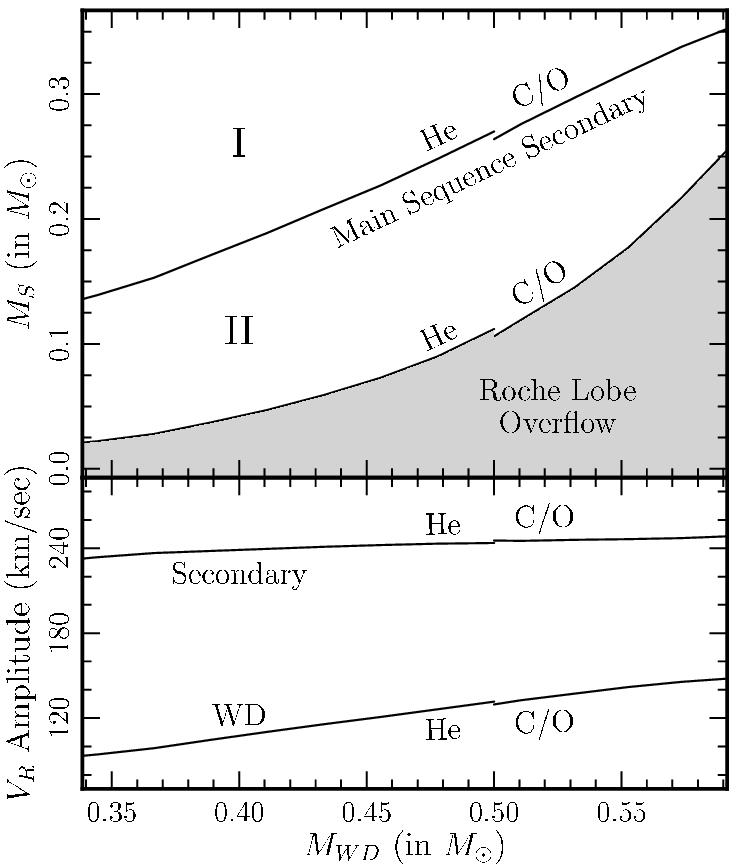}
	\caption{Upper Panel:  Constraints within the $M_S-M_{WD}$ parameter space. All curves incorporate a minimized $\chi^2$ fit to eclipse light curve (see \S\ref{sec:ecl}). All WD $M-R$ relations are derived as follows: He WD from \citet{lga97} and C/O WD from \citet{lga98}. The MS $M-R$ relation is from \citet{ib98}.  The RL boundary is defined using \citet{ppe83}. Region I requires a secondary which is denser than a MS star of its mass, a likely sign of He enrichment.  Region II requires a secondary which is less dense than a MS star of its mass, most likely not in thermal equilibrium and still contracting to the MS.  Lower Panel:  Radial velocity amplitudes for the WD and secondary for the MS solutions in the Upper Panel.}
	\label{fig:msmw}
\end{figure}

When modeling the limb darkening of a WD atmosphere with a linear law, the coefficient depends upon the effective temperature and the bandpass of the filter.  Insufficient data exists to define this dependence, so we used two values bracketing the measured range of similar systems: $u_{LD}=$ 0.3-0.5 \citep{spl06,spl07,pflm07}.

\begin{deluxetable}{ccccccc}
\tablewidth{0pt}
\tablecaption{Best Fit Main Sequence Secondary Solutions\label{tbl:mss}}
\tablehead{
\colhead{$M_{WD}$} & \colhead{$u_{LD}$} & \colhead{$M_S$} & \colhead{$R_S$} & \colhead{$i$} & \colhead{$V_{R,WD}$} & \colhead{$V_{R,S}$} \\
\colhead{($M_{\odot}$)} & \colhead{ } & \colhead{($M_{\odot}$)} & \colhead{($R_{\odot}$)} & \colhead{(deg)} & \colhead{(km/sec)} & \colhead{(km/sec)}  }
\tablecolumns{4}
\scriptsize
\startdata
	0.35 & 0.3 & 0.143 & 0.167 & 79.3 & 95.7 & 234\\
	0.35 & 0.5 & 0.148 & 0.172 & 79.0 & 98.3 & 233\\
	0.58 & 0.3 & 0.340 & 0.320 & 72.4 & 146 & 248\\
	0.58 & 0.5 & 0.349 & 0.327 & 72.0 & 148 & 246\\
\enddata
\end{deluxetable}

Given the present uncertainties and range of RL fillings, we simply assume that the secondary star has the $M_S-R_S$ relation from \citet{ib98}.  The MS solutions for the two extreme accepted WD masses and two limb darkening coefficients are in Table \ref{tbl:mss}.  Further, we investigated how the secondary star solution would change with the WD mass.  In Figure \ref{fig:msmw} we plot the best fit MS star solutions using He core and C/O core WD models from \citet{lga97,lga98} in the $M_S-M_{WD}$ parameter space.

We bound this space for low $M_S$ by requiring that our system not be in RL overflow.  If we assume our system to be at the boundary of RL overflow we can minimize $\chi^2_{\rm fit}$ along the RL overflow boundary shown in Figure \ref{fig:rmch}.  The result of this minimization can be seen in Figure \ref{fig:msmw} as the boundary of the shaded region.  The scaling of this curve is simply understood by considering the geometry of inclination and the circular orbits of two masses.  By requiring an inclination where the WD is just fully eclipsed at zero phase, using the period and transit time of the binary, the radius of the secondary is found as a function of $M_{WD}$ and $M_S$.  Equating this radius with the RL filling radius of \citet{ppe83} gives $M_S$ as a function of $M_{WD}$ whose scaling closely follows that of the curve in Figure \ref{fig:msmw}.  Any system below this line must be in RL overflow.  Figure \ref{fig:msmw} clearly illustrates two regimes for the secondary star.  Region I requires it to be more dense than that of a MS star of the same mass, therefore, if found in this region it would be a likely sign of He enrichment in the secondary from a prior stage of mass transfer \citep{ehpp88}.  Region II requires the secondary star to be less dense than that of a MS star of the same mass, therefore, if found in this region, the secondary would likely not be in thermal equilibrium.  Both of these conditions are only possible if J1435 is a mass transfer system that is temporarily out of contact.  That being said, the most likely explanation for J1435 is a pre-Cataclysmic Variable (pre-CV) where the M dwarf is a MS star that has yet to make contact.


\section{Discussion}
\label{sec:dis}

Our analysis of J1435 is far from complete.  With a single color light curve of the primary eclipse, only broad constraints may be placed upon the parameters of the system (see \S\ref{sec:ecl}).  Therefore, we look to future observations in multiple colors as well as full spectra over the entire orbit to increase the precision and accuracy of the system parameters.  Our observations in a very blue bandpass, while ideal for a ZZ Ceti pulsation search, puts the effects of an M dwarf secondary eclipse at a level far below the precision of our observations.  A redder bandpass would allow for better observation of the secondary eclipse and resulting model fits would provide additional constraint to J1435's components.  Full spectral coverage over the entire orbital period would ideally reveal a double-lined binary system yielding velocity curves for both components.  With these velocity curves, good measurement of the inclination of the orbit can be made.

One spectrum of J1435 was observed in the SDSS \citep{dgy00} and within the errors of our ephemeris the entire exposure was taken out of eclipse.  Visual inspection of the SDSS spectrum and comparison with M dwarf spectra from \citet{jjb07} clearly place the secondary star's spectral type to be later than M4, consistent with under-filling the RL at $P=3$ hrs with $M_S<0.2 M_{\odot}$ \citep{ck06}.  Using the empirical M dwarf colors from \citet{jjb07} we fix the colors of the secondary at $u'-g'>2.3$ and $g'-i'>3.0$.  Therefore, the M dwarf contribution in the BG-39 filter would be $\lesssim4 \%$ of the total system flux.  This is consistent with the $10 \%$ remaining flux observed at maximum eclipse and the lack of an observed secondary eclipse.

Further constraint of the secondary mass is possible by using the SDSS spectrum, synthetic WD spectra, and WD and M dwarf absolute magnitudes.  We used a synthetic WD spectrum provided by G\"{a}nsicke et al. (priv. com., \citealt{arm07}) within the J1435 $T_{\rm eff}$ and $\log g$ parameters and scaled it to fit the Balmer absorption lines shorter than 500 nm in the SDSS spectrum.  In the Johnson $V$ filter at 560 nm, the contribution from a WD with our synthetic spectrum requires the M dwarf to be $\approx2.3$ magnitudes fainter.  Using \citet{jbh06} for WD and \citet{ck06} for M dwarf absolute magnitudes, we find the M dwarf to be in the spectral range M4-M6 with $M_S=0.11-0.20 M_{\odot}$.  This analysis is in contrast to that of \citet{nms06} which found the M dwarf spectral type of J1435 to be M2-M4 with $M_S=0.20-0.55 M_{\odot}$.  It is clear that much more rigorous spectral analysis needs to be carried out if it is to be used to constrain the secondary mass.

We expect that J1435 is a pre-CV system with the M dwarf secondary on or near the MS.  Two systems with similar parameters to J1435, MS Peg and NN Ser, have calculated times to contact of $\approx1.5$ Gyrs \citep{mrs03}.  There are little more than a dozen eclipsing double-lined spectroscopic binaries containing M dwarfs below $1 M_{\odot}$ \citep{ajb06,lh06,ir06,tby06,jd07,mlm07,jss07}. If J1435 were found to be a double-lined spectroscopic binary, from which velocity curves for both components could be found and its parameters found to high precision, it would be a significant addition to this group especially being below $0.35 M_{\odot}$, where theory \citep{ls97,ib98,sy01} would benefit.

\acknowledgments
 
We thank both referees for comments that clarified our presentation.  This work was supported by the National Science Foundation under grants PHY 05-51164 and AST 07-07633.


\end{document}